\magnification\magstep1
\font\BBig=cmr10 scaled\magstep2


\def\title{
{\bf\BBig
\centerline{An integrable}
\bigskip
\centerline{time-dependent non-linear Schr\"odinger equation}
\bigskip
}
} 


\def\authors{
\centerline{
P.~A.~HORV\'ATHY\foot{e-mail: horvathy@univ-tours.fr}
 and
J.-C.~YERA\foot{e-mail: yera@univ-tours.fr}}
\bigskip
\centerline{
D\'epartement de Math\'ematiques}
\medskip
\centerline{Universit\'e de Tours}
\medskip
\centerline{Parc de Grandmont,
F--37200 TOURS (France)
}
}

\def\runningauthors{
Horv\'athy \&
Yera
}

\def\runningtitle{
An integrable 
 time-dependent non-linear Schr\"odinger equation 
}


\voffset = 1cm 
\baselineskip = 14pt 

\headline ={
\ifnum\pageno=1\hfill
\else\ifodd\pageno\hfil\tenit\runningtitle\hfil\tenrm\folio
\else\tenrm\folio\hfil\tenit\runningauthors\hfil
\fi
\fi}

\nopagenumbers
\footline={\hfil} 


\def\and{\qquad\hbox{and}\qquad}

\def\kikezd{\parag\underbar} 

\def\IR{{\bf R}}
\def\smallover#1/#2{\hbox{$\textstyle{#1\over#2}$}}
\def\2{{\smallover 1/2}}
\def\ccr{\cr\noalign{\medskip}} 
\def\parag{\hfil\break} 
\def\={\!=\!}

\def\const{{\rm const}}


\newcount\ch 
\newcount\eq 
\newcount\foo 
\newcount\ref 

\def\chapter#1{
\parag\eq = 1\advance\ch by 1{\bf\the\ch.\enskip#1}
}

\def\equation{
\leqno(
\the\eq)\global\advance\eq by 1
}

\def\foot#1{
\footnote{($^{\the\foo}$)}{#1}\advance\foo by 1
} 

\def\reference{
\parag [\number\ref]\ \advance\ref by 1
}

\ch = 1 
\eq = 1
\foo = 1 
\ref = 1 
\pageno=1


\title
\vskip10mm
\authors
\vskip.20in

\parag{\bf Abstract.}
{\it The cubic non-linear Schr\"odinger equation 
(NLS), where the coefficient of the non-linear term 
can be a function $F(t,x)$,
is shown to pass the 
Painlev\'e test of Weiss, Tabor, and Carnevale
only for 
$F=(a+bt)^{-1}$, where $a$ and $b$ constants.
This is explained by transforming the time-dependent system into the 
ordinary NLS (with $F=\const$.) by means of a time-dependent
non-linear transformation, related to the conformal
properties of non-relativistic space-time.
}

\vskip15mm
\noindent
(\the\day/\the\month/\the\year)
\bigskip
\medskip\noindent
{\sl Physics Letter} {\bf A} \ (submitted).
\vskip35mm

\noindent
\vskip5mm
\vfill\eject

Let us consider the cubic 
non-linear Schr\"odinger equation (NLS),
$$
iu_t+u_{xx}+F(t,x)\vert u\vert^2u=0,
\equation
$$ 
where $u=u(t,x)$ is a complex function in $1+1$ space-time dimension.
When the coefficient $F(t,x)$ of the non-linearity is a
constant, this is the usual NLS, which is known 
to be integrable. But what happens, when 
 the coefficient 
$F(t,x)$ is a {\it function} rather then just a constant~?
Performing the
Painlev\'e analysis of Weiss, Tabor and Carnevale
[1],
we show 

\kikezd{Theorem1}~:
{\it The generalized non-linear Schr\"odinger equation (1)
only passes the Painlev\'e test if the coefficient of the non-linear term
is of the form} 
$$
F(t,x)={1\over at+b}, 
\qquad
a, b=\const.
\equation
$$
 
\kikezd{Proof}.
As it is usual in studying non-linear Schr\"odiger-type equations
[2], we consider Eqn. (1) together with its complex conjugate ($v=u^*$),
$$
\eqalign{
 iu_t+u_{xx}+Fu^2v&=0,
\cr
-iv_t+v_{xx}+Fv^2u&=0.
\cr}
\equation
$$

This system will pass the Painlev\'e  test if $u$ et $v$ 
have generalised Laurent series expansions,
$$
u=\sum_{n=0}^{+\infty}u_n\xi^{n-p},\qquad
v=\sum_{n=0}^{+\infty}v_n\xi^{n-q},
\equation
$$
($u_n\equiv u_n(x,t)$, $v_n\equiv v_n(x,t)$ and $\xi\equiv\xi(x,t)$)
in the neighbourhood of the singular manifold $\xi(x,t)=0$, $\xi_x\not=0$,
with a sufficient number of free
coefficients. Using the results of Tabor, and of Weiss [3],
it is enough to consider
$
\xi=x+\psi(t);
$
then $u_n$ and $v_n$ become functions de $t$ alone,
$u_{n}\equiv u_{n}(t),\
v_{n}\equiv v_{n}(t). 
$ 
Checking the dominant terms, 
$u\sim u_0\xi^{-p}$, $v\sim v_0\xi^{-q}$, 
using the above remark, we get
$$
p=q=1,
\qquad
F\,u_0v_0=-2.
\equation
$$
Hence  $F$ can only depend on $t$.
Now inserting the developments (4) of $u$ and $v$ into (3),
the terms in $\xi^k$, $k\geq -3$ read
$$
\eqalign{
& i\Big(u_{k+1,t}+(k+1)u_{k+2}\xi_t\Big)+(k+2)(k+1)u_{k+3}
 +F\Big(\sum_{i+j+l=k+3}u_iu_jv_l\Big)=0,
 \cr
& i\Big(v_{k+1,t}+(k+1)v_{k+2}\xi_t\Big)+(k+2)(k+1)v_{k+3}
 +F\Big(\sum_{i+j+l=k+3}v_iv_ju_l\Big)=0.
 \cr}
\equation
$$
(The condition  (5) is recovered for $k=-3$).    
The coefficients $u_n$, $v_n$ of the series (3) are
given by the system $S_n$ ($k=n-3$),
$$
\eqalign{
&[(n-1)(n-2)-4]u_n+Fu_0^2v_n=A_n,
\cr
& Fv_0^2u_n+[(n-1)(n-2)-4]v_n=B_n,
\cr}
\equation
$$
where $A_n$ et $B_n$ only contain those terms $u_i$, $v_j$ with
$i,j<n$. The determinant of the system is 
$$
\det S_n=n(n-4)(n-3)(n+1).
\equation
$$
Then (3) passes the Painlev\'e test if, for each $n=0,3,4$, 
one of the coefficients $u_n$, $v_n$ can be arbitrary. 
For $n=0$, (5) implies 
that this is indeed true either
for  $u_0$ or $v_0$. 
For $n=1$ and $n=2$, the system (6)-(7) is readily solved, yielding 
$$\eqalign{
&u_1=-{i\over2}u_0\xi_t,
\qquad
v_1={i\over2}v_0\xi_t,
\cr
&6v_0u_2=iv_{0,t}u_0+2iu_{0,t}v_0-{1\over2}u_0v_0(\xi_t)^2,
\cr
&6u_0v_2=-iu_{0,t}v_0-2iv_{0,t}u_0-{1\over2}u_0v_0(\xi_t)^2.
\cr}
\equation
$$

$n=3$ has to be a resonance; using condition (5),
the system (7) becomes
$$
\eqalign{
&-2v_0u_3-2u_0v_3=A_3v_0,
\cr
&-2v_0u_3-2u_0v_3=B_3u_0,
\cr}
$$
which requires $A_3v_0=B_3u_0$. 
But using the expressions of $A_{3}$ and $B_3$, with the
help of ``Mathematica'' we find 
$$
2FA_3=u_0(F_t\xi_t-F\xi_{tt}),\qquad u_0F^2B_3=F\xi_{tt}-F_t\xi_t,
$$
so that the required condition indeed holds.

$n=4$ has also to be a resonance;  we find, as before,
$$
\eqalign{
2v_0u_4-2u_0v_4&=A_4v_0,
\cr
-2v_0u_4-2u_0v_4&=B_4u_0,
\cr}
$$
enforcing the relation
$
v_0A_4=-u_0B_4.
$ 
Now using the expressions of
$v_0$, $u_1$, $v_1$, $u_2$, $v_2$ as functions of $u_0$, $F$,  
$u_3$, $v_3$,  ``Mathematica'' yields
$$
\eqalign{
6u_0F^2A_4=
&\Big(-F^2u_{0,t}^2-2iu_0^2F^2\xi_t\xi_{tt}+u_0F^2u_{0,tt}
+iu_0^2F\xi_t^2F_t-u_0Fu_{0,t}F_t
\cr
&+2u_0F_t^2-u_0^2FF_{tt}\Big),
\cr
3u_0^3F^3B_4=
&\Big(-F^2u_{0,t}^2-2iu_0^2F^2\xi_t\xi_{tt}+u_0F^2u_{0,tt}
+iu_0^2F\xi_t^2F_t-u_0Fu_{0,t}F_t
\cr
&-4u_0F_t^2+2u_0^2FF_{tt}\Big).
\cr}
$$
Then our constraint implies that $2F_t^2-FF_{tt}=0$, i. e.
$
{d^2\over dt^2}\big({1\over F}\Big)=0.
$ 
Hence $F^{-1}(x,t)=at+b$, as stated. 
\goodbreak

For $a=0$,
$
F(t,x)
$
in Eqn. (1) is a constant,
and we recover the usual NLS with its known solutions.
For $a\neq0$, the equation becomes explicitly time-dependent.
Assuming,
for simplicity, that $a=1$ and $b=0$, it reads
$$
iu_{t}+u_{xx}+{1\over t}\vert u\vert^2u=0.
\equation
$$

This equation can also be solved. Generalizing the usual
``travelling soliton'', let us seek, for example,
 a solution of the form
$$
u(t,x)=
e^{i(x^2/4t-1/t)}\,f(t,x),
\equation
$$
where $f(t,x)$ is some real function. Inserting the  Ansatz
(11) into (10), the real and imaginary parts yield
$$
\eqalign{
&f_{xx}-{1\over t^2}f+{1\over t}f^3=0,
\cr
&f_{t}+{x\over t}f_{x}+{1\over2t}f=0.
\cr}
\equation
$$

Time dependence can now be eliminated~:
setting $f(t,x)=t^{-1/2}g(-1/t,-x/t)$ transforms (12) into
$$
\eqalign{
&g_{xx}-g+g^3=0,
\cr
&g_{t}=0.
\cr}
\equation
$$

Multiplying the first equation by
$g_{x}$ yields a spatial divergence;
then requiring the asymptotic behaviour
$g(t,\pm\infty)=0=g_{t}(t,\pm\infty)$ 
and taking into account the second equation 
yields
$
g(t,x)={\sqrt{2}/{\rm sech}[x-x_{0}]}.
$
In conclusion, we find the soliton
$$
u(t,x)={e^{i(x^2/4t-1/t)}\,\over\sqrt{t}}\,
{\sqrt{2}\over\cosh\big[-x/t-x_{0}\big]}.
\equation
$$

 It is worth pointing out that the eqns. (13) are essentially the same as
those met when constructing travelling solitons for the ordinary NLS ---
and this is not a pure coincidence.
We have in fact

\kikezd{Theorem2}.
$$
u(t,x)={1\over\sqrt{t}}\exp\Big[{ix^2\over4t}\Big]\,
\psi\big(-{1/t},-{x/t}\big)
\equation
$$
{\it satisfies the time-dependent equation (10)
if and only if 
$\psi(t,x)$ solves Eqn. (1) with $F=1$}.
\goodbreak
\vskip2mm
This can readily be proved by a direct calculation. Inserting (15) 
into (10), we find in fact
$$
t^{-5/2}\exp\Big[{ix^2\over4t}\Big]\,\bigg(
i\psi_{t}+\psi_{xx}+\vert\psi\vert^2\psi\bigg).
$$

Our soliton (14) constructed above comes in fact from the
well-known ``standing soliton'' solution of the NLS,
$$
\psi_{0}(t,x)={\sqrt{2}\,e^{it}\over\cosh[x-x_{0}]},
\equation
$$
by the transformation (15). More general solutions
could be obtained starting with the ``travelling soliton''
$$
\psi(t,x)=e^{i(vt-kx)}{\sqrt{2}\,a\over\cosh[a(x+kt)]},
\qquad
a=\sqrt{k^2+v}.
\equation
$$

Where does the formula (15) come from~?
To explain it, let us remember 
that the non-linear space-time transformation
$$
D:\pmatrix{ t\cr x\cr}
\to
\pmatrix{-\displaystyle{1/t}\cr -\displaystyle{x/t}\cr}
\equation
$$
has already been met in a rather different context, namely
in describing planetary motion when the gravitational ``constant''
changes inversely with time, as suggested by Dirac [4].
One shows in fact that 
$
\vec{r}(t)=t\,\vec{r}^*\big(-{1/t})
$ 
describes  planetary 
motion with Newton's ``constant'' varying as $G(t)=G_0/t$, whenever
$\vec{r}^*(t)$ describes ordinary planetary motion, i.e. the one
with a constant gravitational constant, $G(t)=G_0$ [5].

The strange-looking transformation (18) is indeed related to the
conformal structure of non-relativistic space-time [6].
It has been noticed in fact almost thirty years ago, that the space-time
transformations
$$\matrix{
\pmatrix{t\cr x\cr}
\to
\pmatrix{t'\cr x'\cr}=
\pmatrix{ 
\displaystyle{\delta^2}t\cr
\displaystyle{\delta}\, x\cr},
\qquad\hfill
&0\neq\delta\in\IR\hfill
&\hbox{dilatations}\hfill
\ccr
\pmatrix{t\cr x\cr}
\to
\pmatrix{t'\cr x'\cr}=
\pmatrix{
\displaystyle{t\over1-\kappa t}\ccr
\displaystyle{x\over1-\kappa t}\cr},
\qquad\hfill
&1\neq\kappa\in\IR\hfill
&\hbox{expansions}\hfill
\ccr
\pmatrix{t\cr x\cr}
\to
\pmatrix{t'\cr x'\cr}=
\pmatrix{t+\epsilon\cr x\cr},
\qquad\hfill
&\epsilon\in\IR\hfill
&\hbox{time translations}\hfill
\cr}
\equation
$$
implemented on wave functions according to
$$
u'(t',x')=\left\{
\matrix{
&\displaystyle{1\over\delta}u(t,x)\hfill
\ccr
&(1-\kappa t)\exp
\Big[i\displaystyle{\kappa x^2\over4(1-\kappa t)}\Big]u(t,x)
\hfill
\ccr
&u(t,x)
\hfill
\cr}\right.
\equation
$$
permute the solutions of
 the free Schr\"odinger equation [7]. 
In other words, they are {\it symmetries} for the free Schr\"odinger 
equation. (The generators in (19) span in fact an ${\rm SL}(2,\IR)$ 
group; when added to the obvious galilean symmetry, 
the so-called Schr\"odinger group is obtained.
A Dirac monopole, an Aharonov-Bohm 
vector potential, and an inverse-square potential can also be included).

The  transformation $D$ in Eqn. (18) belongs to
this symmetry group: it is in fact 
 (i)  a time translation with 
$\epsilon=1$, (ii) followed an expansion with 
$\kappa=1$, (iii) followed by a second time-translation with
$\epsilon=1$.
It is hence a symmetry for the free (linear) Schr\"odinger
equation. Its action on $\psi$, deduced from (20), is precisely (15).

The cubic NLS with non-linearity $F=\const$.
 is no more $SL(2,\IR)$ invariant\foot{
 Galilean symmetry can be used to produce further
 solutions --- just like the ``travelling soliton'' (17) can be 
 obtained from the ``standing one'' in (16) by a galilean boost.
Full Schr\"odinger invariance yielding expanded and dilated solutions 
can be restored by replacing the cubic non-linear term by
the fifth-order non-linearity $\vert\psi\vert^4\psi$.}.
In particular, the 
 transformation $D$ in (18), implemented as in Eq. (15)
 carries the cubic term into the
time-dependent term
$(1/t)\vert u\vert^2u$ --- just like Newton's gravitational potential
$G_{0}/r$ with $G_0=\const$. is carried into the time-dependent
Dirac expression $t^{-1}G_{0}/r$ [5].

Our results should be compared with the 
those of Chen et al. [8], who prove that the equation
$
iu_t+u_{xx}+F(\vert u\vert^2)u=0
$
can be solved by inverse scattering
if and only if
$
F(\vert u\vert^2)=\lambda\vert u\vert^2,
$
where $\lambda=\const$. Note, however, that Chen et al. only
study the case when the functional $F(\vert u\vert^2)$ 
is independent of the space-time coordinates $t$ and $x$.

In this Letter, we only studied the case of 
$d=1$ space dimension. 
Similar results would hold for any $d\geq1$, though.

\goodbreak

\goodbreak
\kikezd{Acknowledgements}.
J.-C. Y. acknowledges
the {\it Laboratoire de Math\'emathiques et de Phy\-si\-que Th\'eo\-ri\-que}
of Tours University for hospitality, and the 
{\it Gouvernement de La C\^ote d'Ivoire}
for a doctoral scholarship. 
We are indebted to Mokhtar Hassa\"\i ne for discussions.

\goodbreak
\vskip5mm\goodbreak
\centerline{\bf References}

\reference 
J. Weiss, M. Tabor, and G. Carnevale,
{\sl J. Math. Phys}. {\bf 24}, 522 (1983).

\reference 
J. Weiss, 
{\sl J. Math. Phys}. {\bf 26},  258 (1985).

\reference 
J. M. Tabor,
{\sl Chaos and integrability in non-linear dynamics},
Wiley, (1989) see also
J. Weiss, 
{\sl J. Math. Phys}. {\bf 24},  1405 (1983).

\reference 
P.~A.~M. Dirac, {\sl Proc. R. Soc}. {\bf A165}, 199 (1938).

\reference 
J. P. Vinti {\sl Mon. Not. R. Astron. Soc}.
{\bf 169}, 417 (1974);
see also
C.~Duval, G.~Gibbons and P.~A. Horv\'athy,
{\sl Phys. Rev}.  {\bf D43}, 3907 (1991).

\reference 
C.~Duval, G.~Burdet, H.~P. K\"unzle and M.~Perrin,
{\sl Phys. Rev}. {\bf D31}, 1841 (1985).

\reference 
R.~Jackiw, {\sl Physics Today}, {\bf 25}, 23 (1972);
U. Niederer, {\sl Helv. Phys. Acta} {\bf 45}, 802 (1972);
C.~R. Hagen, {\sl Phys. Rev}. {\bf D5}, 377 (1972).

\reference 
H. H. Chen, Y. C. Lee et C. S. Liu, 
{\sl Physica Scripta} {\bf 20}, 490 (1979).


\end